\documentclass[apl, aip, reprint]{revtex4-1}
\usepackage{CJK}
\usepackage{graphicx}
\usepackage{amsmath}
\usepackage{amssymb}
\usepackage{color}
\usepackage[english]{babel}
\begin{document}

\title{Enhanced Rashba spin-orbit coupling in core-shell nanowires by the interfacial effect}

\author{Pawe{\l} W\'ojcik}
\email{pawel.wojcik@fis.agh.edu.pl}
\affiliation{AGH University of Science and Technology, Faculty of
Physics and Applied Computer Science, 30-059 Krakow, Al. Mickiewicza 30, Poland}

\author{Andrea Bertoni}
\email{andrea.bertoni@nano.cnr.it}
\affiliation{S3, Istituto Nanoscienze-CNR, Via Campi 213/a, 41125 Modena, Italy}

\author{Guido Goldoni}
\email{guido.goldoni@unimore.it}
\affiliation{Department of Physics, Informatics and Mathematics, University od Modena and Reggio Emilia, Italy}
\affiliation{S3, Istituto Nanoscienze-CNR, Via Campi 213/a, 41125 Modena, Italy}

\date{\today}

\begin{abstract}
We report on $\vec{k}\cdot\vec{p}$ calculations of Rashba spin-orbit coupling controlled by external gates in InAs/InAsP core-shell nanowires. We show that charge spilling in the barrier material 
allows for a stronger symmetry breaking than in homoegenous nano-materials, inducing a specific interface-related contribution to spin-orbit 
coupling. Our results qualitatively agree with recent experiments [S. Futhemeier \textit{et al.}, Nat. Commun. \textbf{7}, 12413 (2016)] and suggest additional wavefunction engineering strategies to 
enhance and control spin-orbit coupling.  
\end{abstract}

\maketitle

Understanding and controlling spin-orbit coupling (SOC) is critical in semiconductor physics. In particular, in semiconductor nanowires 
(NWs)~\cite{Kokurin2015,Kokurin2014,Kammermeier2016,Winkler2017,Wojcik2018,Campos, Luo} SOC is essential for to the development of a suitable hardware for topological quantum 
computation~\cite{Alicea,Sau2017}, with qubits encoded in zero-mode Majorana states which are supported by hybrid semiconductor-superconductor NWs~\cite{Kitaev2003, Sau2012, Mourik2012, 
Mastomaki2017NRes, Albrecht2016, Manolescu2017}. Among other parameters, qubit protection at sufficiently high temperatures relies on a large SOC which determines the topological gap. 
Additionally, electrical control of SOC is necessary in the realization of spintronic 
devices~\cite{Fabian2007,Datta,Schliemann,Wojcik2016,Wojcik2014,Wojcik2017,Das2012,Debray2009,Kohda2012,RossellaNN2014,Iorio2018arXiv}.  

SOC arises from the absence of inversion symmetry of the electrostatic potential. In semiconductor NWs, typically having a prismatic shape, finite SOC may be induced by distorting the quantum 
confinement (Rashba SOC~\cite{Rashba}) by means of external gates, with the advantage of electrical control. A lattice contribution (Dresselhaus SOC~\cite{Dresselhaus}) is typically small and may 
vanish in specific crystallographic directions - for zincblende NWs, the Dresselhaus term vanishes along [111] due to the inversion symmetry. 

The Rashba SOC constant $\alpha_R$ has been investigated experimentally in homogeneous NWs based on the strong SOC materials InSb~\cite{vanWeperen2015,Kammhuber2017} and 
InAs~\cite{Scherubl2016,Hansen2005,Dhara2009,Roulleau2010,Estevez2010}. Recently~\cite{Wojcik2018}, we reported on a $\vec{k}\cdot\vec{p}$ approach applied to \textit{homogeneous} NWs which predicts 
$\alpha_R$ from compositional and structural parameters only. Our calculations performed for InSb NWs~\cite{Wojcik2018} and InAs NWs~\cite{WojcikUNP} generally confirm recent experiments in 
homogeneous NWs~\cite{vanWeperen2015,Scherubl2016}, exposing values of $\alpha_R$ exceeding by one order of magnitude those reported for 2D analogous planar 
systems~\cite{Kallaher2010,Kallaher2010b,Herling}. Moreover, $\alpha_R$ proved to be strongly tunable with external gates in samples and configurations which can be routinely realized with current 
technology. 

For a quantitative prediction of SOC, it is necessary to take into account valence-to-conduction band coupling, the explicit geometry and crystal structure of the NW, and the electron gas 
distribution which, in turn, must be self-consistently determined by quantum confinement effects, interaction with dopants and electron-electron interaction. Indeed, in NWs the electron gas 
localization, and ensuing SOC, is a non-trivial result of competing energy contributions. As a function of doping concentration and ensuing free charge density, the electron gas evolves from a broad 
cylindrical distribution in the NW core (low density regime) to coupled quasi-1D and quasi-2D channels at the NW edges (large density 
regime)~\cite{Bertoni2011,FunkNL2013,JadczakNL2014,RoyoPRB2014}. 
Until  polygonal symmetry holds, $\alpha_R =0 $ regardless. 
However, external gates easily remove the symmetry; again, how $\alpha_R$ moves from zero under the influence of the external gates strongly depends on the charge density regime~\cite{Wojcik2018}.
\begin{figure}[!b]
	\begin{center}
		\includegraphics[scale=0.2]{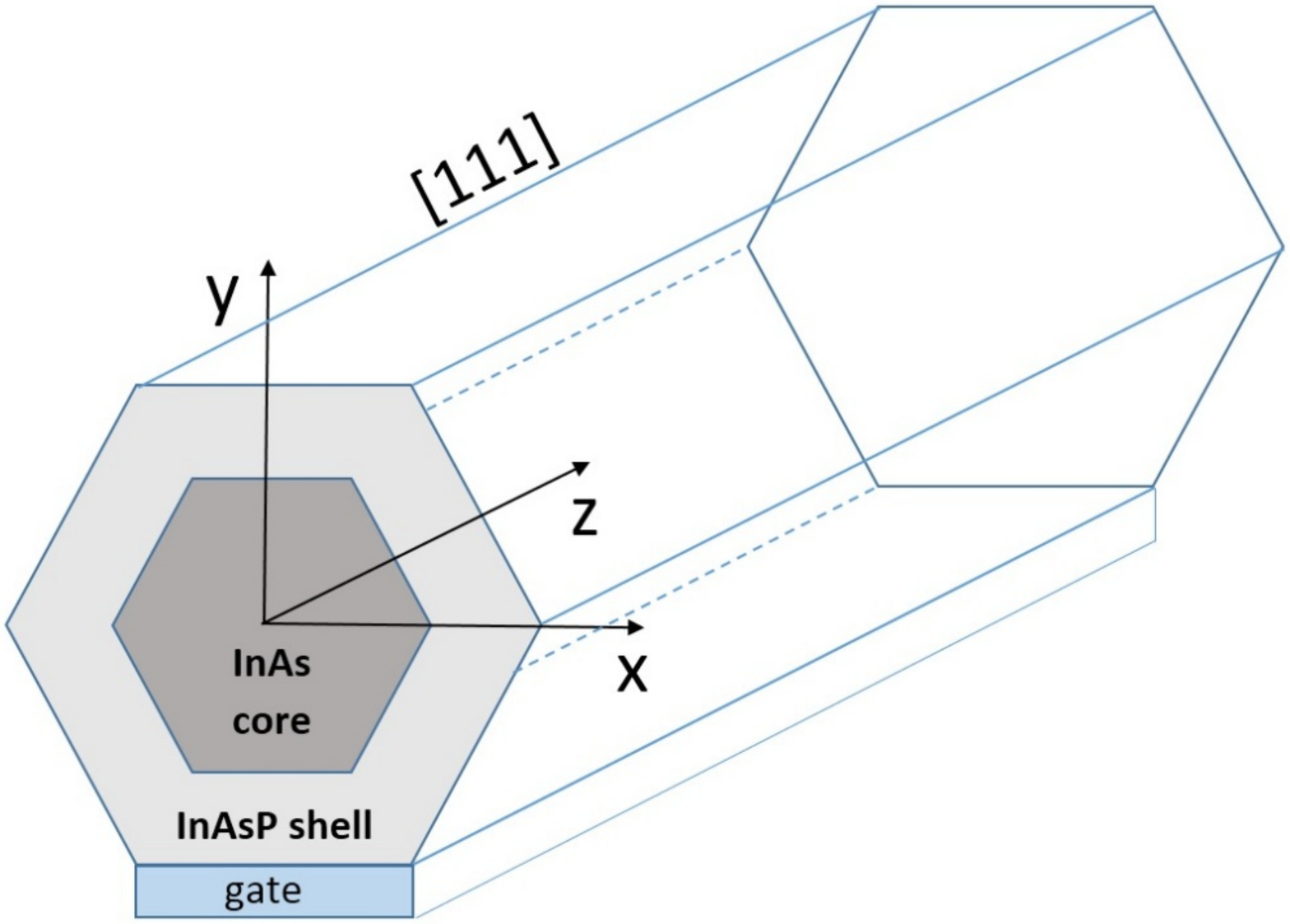}
		\caption{Schematics of a InAs/InAsP CSNW grown along [111] with a bottom gate.}
		\label{fig1}
	\end{center}
\end{figure}

In this Letter we extend and apply the $\vec{k}\cdot\vec{p}$ approach to core-shell NWs (CSNWs) and expose a novel mechanism through which SOC can be further tailored, and possibly 
enhanced. Epitaxially overgrown shells are often used in NW technology, either as a passivating layer improving optical performance~\cite{Jabeen2008}, or as a technique to engineer radial 
heterostructures~\cite{Spirkoska2009}. Here we show that CSNWs allow for an increased flexibility in distorting the electron gas of the NWs, giving rise to a specific, interfacial SOC 
contribution~\cite{Devizorova2013,Ivchenk1996} which substantially increases the total SOC. 
We make the case for InAs/InAsP CSNWs, a systems of specific interest in photonics~\cite{Treu2013} and electrical engineering~\cite{Liu2013}. 
Our results qualitatively agree with the recent experiments by Furthmeier \textit{et al.} in Ref.~\onlinecite{Furthmeier}, where the enhancement of SO coupling was measured in GaAs/AlGaAs CSNW, and 
establish a strategy to increase the SOC in Majorana InAs NWs.

We consider CSNWs with hexagonal cross-section~\cite{Sitek2018} grown along [111] (see Fig.~\ref{fig1}), assuming in-wire translational invariance along $z$. 
The used $\vec{k}\cdot\vec{p}$ approach is described in full in Ref.~\onlinecite{Wojcik2018}; here we focus on generalizations required to account for the contribution of the internal heterointerface. 
The $8 \times 8$ Kane Hamiltonian is given by~\cite{Fabian2007} 
\begin{equation}
\label{eq:KH}
H_{8\times 8}=\left ( 
\begin{array}{cc}
H_c & H_{cv} \\
H^{\dagger}_{cv} & H_v
\end{array}
\right ),
\end{equation}
where $H_c$ and $H_v$ are the diagonal matrices corresponding to the conduction ($\Gamma _{6c}$) and valence ($\Gamma _{8v}$, $\Gamma _{7v}$) bands whose expressions are given in Ref.~\onlinecite{Wojcik2018}. 
Using the perturbative transformation $\mathcal{H}(E)=H_c+H_{cv}(H_v-E)^{-1}H_{cv}^{\dagger}$, the Hamiltonian (\ref{eq:KH}) reduces to a $2\times 2$ effective Hamiltonian for the conduction band electrons. 
Emphasizing the dependence of material parameters on the position, $\vec{\rho}=(x,y)$, 
\begin{multline}
\label{eq:3DH}
 \mathcal{H} = \left [ -\frac{\hbar^2}{2} \nabla _{2D} \frac{1}{m^*(\vec{\rho})} \nabla _{2D} 
                       + \frac{\hbar^2 k_z^2}{2m^*(\vec{\rho})}
                       + E_c(\vec{\rho}) + V(\vec{\rho}) \right ]  \\
\times\mathbf{1}_{2\times 2}
 + \left[\hat{\alpha}_x(\vec{\rho}) \sigma _x + \hat{\alpha}_y(\vec{\rho}) \sigma _y\right]k_z,
\end{multline}
where $\sigma _{x(y)}$ are the Pauli matrices and $m^*$ is the effective mass given by
\begin{equation}
 \frac{1}{m^*(\vec{\rho})}=\frac{1}{m_0}+\frac{2P^2}{3\hbar ^2} \left ( \frac{2}{E_0(\vec{\rho})} + \frac{1}{E_0(\vec{\rho})+\Delta_0(\vec{\rho})}
\right ),
\end{equation}
where $P$ is the conduction-to-valance band coupling parameter. \\
In Eq.~(\ref{eq:3DH}), $\hat{\alpha}_x$, $\hat{\alpha}_y$ are the SOC operators
\begin{eqnarray}
\label{ax}
 \hat{\alpha} _x &=&  \frac{i}{3} P^2  \hat{k}_y \beta(\vec{\rho}) - \frac{i}{3} P^2 \beta(\vec{\rho}) \hat{k}_y,  \\
 \label{ay}
\hat{\alpha} _y &=&  \frac{i}{3} P^2  \hat{k}_x \beta(\vec{\rho}) - \frac{i}{3} P^2 \beta(\vec{\rho}) \hat{k}_x,
\end{eqnarray}
and $\beta(\vec{\rho})$ is a material-dependent coefficient obtained as follows.
In the $i$-th layer
\begin{eqnarray}
\label{beta}
\beta_i(\vec{\rho}) &=&  \frac{1}{E_{c,i}+V(\vec{\rho})-E_{0,i}-E}  \\ 
&-& \frac{1}{E_{c,i}+V(\vec{\rho})-E_{0,i}-\Delta _{0,i}-E} , \nonumber
\end{eqnarray}
where $E_c$, $E_0$ and $\Delta_0$ are the conduction band edge, the energy gap and the split-off band gap, respectively. 
Assuming that the above parameters change as a step-like function at the interfaces
\begin{equation}
 \beta(\vec{\rho}) = 
 \sum _i [\beta _i(\vec{\rho}) - \beta _{i+1}(\vec{\rho})] \Omega_i(\vec{\rho}),
\end{equation}
where the sum is carried out over all the layers,
and $\Omega_i(\vec{\rho})$ is the shape function, which for the hexagonal section is given by
\begin{eqnarray}
\label{eq:omega}
 \Omega _i (\vec{\rho})&=& 
 [\theta(x+x_{i})-\theta(x-x_i)][\theta(y+y_{i})-\theta(y-y_i)] \nonumber \\
                 &\times& [\theta(x-y+x_{i})-\theta(x-y-x_i)],
\end{eqnarray}
where $\theta$ is the Heaviside'a function and $(x_i,y_i)$ denotes the position of the $(i)$-th interface. Further Taylor expansion gives 
\begin{eqnarray}
\label{taylor}
\beta_i(\vec{\rho}) & \approx &  \left ( \frac{1}{E_{0,i}+\Delta_{0_i}}-\frac{1}{E_{0,i}} \right )   \\
&+& \left (\frac{1}{E_{0,i}^2} - \frac{1}{(E_{0,i}+\Delta_{0,i})^2} \right ) (E_{c,i}+V(x,y)-E). \nonumber
\end{eqnarray}
Substituting (\ref{taylor}) into Eqs.~(\ref{ax}) and (\ref{ay}), the Rashba coupling constants can be written as
\begin{equation}
 \alpha _{x(y)}(\vec{\rho}) = \alpha^{V}_{x(y)}(\vec{\rho}) + \alpha^{int} _{x(y)}(\vec{\rho}) \,,
\end{equation}
\textit{i.e.}, the  sum of the SOC induced by the electrostatic potential asymmetry,
\begin{eqnarray}
\label{eq:aVV}
\alpha _{x(y)} ^V(\vec{\rho}) & \approx &  \sum _i  \frac{1}{3} P^2  \left ( \frac{1}{E_{0,i}^2} - \frac{1}{(E_{0,i}+\Delta_{0,i})^2} \right ) \frac{\partial
V(\vec{\rho})}{\partial y(x)}, \nonumber \\ 
\end{eqnarray}
and the interface SOC, related to the electric field at the interfaces between shells,
\begin{equation}
\label{eq:aix}
\alpha _{x(y)} ^{int}(\vec{\rho}) \approx \sum _i  \frac{1}{3} P^2  \left ( \tilde{\beta} _i - \tilde{\beta} _{i+1} \right ) \frac{\partial
\Omega_i(\vec{\rho})}{\partial y(x)}, 
\end{equation}
with 
$
\tilde{\beta _{i}} = \frac{1}{E_{0,i}+\Delta_{0,i}}-\frac{1}{E_{0,i}}. 
$

Projecting the 3D Hamiltonian (\ref{eq:3DH}) on the basis of in-wire states $\psi_n(\vec{\rho}) \exp(i k_z z)$, where the envelope functions $\psi_n(\vec{\rho})$ are determined by the strong 
confinement in the lateral direction, leads to SOC matrix elements 
\begin{eqnarray}
\alpha_{x(y)}^{\gamma,nm}=\int \int \psi_n(\vec{\rho}) \alpha^{\gamma}_{x(y)}(\vec{\rho}) \psi_m(\vec{\rho}) d\vec{\rho}\,,
\label{eq:agamma}
\end{eqnarray}
where $\gamma$ identifies the electrostatic ($\gamma = V$) or the interfacial ($\gamma = int$) contribution.

For the NW in Fig.~1 with a single bottom gate, $\alpha_{y}^{\gamma,nn} = 0$ due to inversion symmetry about $y$. 
Moreover, here we focus on the lowest intra-subband coefficient, $n=1$.
Below we discuss the SOC constant $\alpha_R=\alpha^{11}_x$ and corresponding interfacial and electrostatic components, $\alpha^\gamma_R=\alpha^{V,11}_x$ and $\alpha^\gamma_R=\alpha^{\mathrm{int},11}_x$, respectively.

The electronic states in the CSNW section, $\psi_n(\vec{\rho})$, are calculated by a mean-field self-consistent Sch{\"o}dinger-Poisson approach~\cite{Bertoni2011}. 
We neglect the exchange-correlation potential which is substantially smaller than the Hartree potential~\cite{Bertoni2011,Nanoletter_XC,Royo_XC}.
The gradient of the self-consistent potential $V(\vec{\rho})$ and the corresponding envelope functions $\psi_n(\vec{\rho})$ are finally used to determine $\alpha_R$ from Eq.~(\ref{eq:agamma}). 

Material parameters mismatch at the interfaces is taken into account solving the eigenproblem $\mathcal{H}\psi_n=E\psi_n$ with boundary conditions~\cite{Silva,Devizorova2013} 
\begin{eqnarray}
\label{eq:bc1}
&& \psi^{(i)}_n(\vec{\rho}_k)=\psi^{(j)}_n(\vec{\rho}_k) \\
\label{eq:bc2}
&& \frac{\hbar ^2}{2m^{*(i)}}\nabla _{2D}\psi^{(i)}_n(\vec{\rho}_k) -\frac{\hbar ^2}{2m^{*(j)}}\nabla _{2D}\psi^{(j)}_n(\vec{\rho}_k) 
\\
&& +[\beta^{(j)}(\vec{\rho}_k)- \beta^{(i)}(\vec{\rho}_k)](\sigma_x+\sigma_y)k_z \psi^{(i)}_n(\vec{\rho}_k)=0, \nonumber 
\end{eqnarray}
where $\vec{\rho}_k$ is the position of the interface between $i$-th and $j$-th shells. Equations (\ref{eq:bc1}), (\ref{eq:bc2}), depend on both the potential $V(\vec{\rho})$ at the interface and the energy $E$. 
We eliminate this dependence neglecting the term proportional to $(E_{c,i}+V(\vec{\rho})-E)$ in the Taylor series, Eq.~(\ref{taylor}). 
Then, the interface contributions (\ref{eq:aix}) are determined fully by material parameters. 
This assumption, justified when $\beta^{(j)}- \beta^{(i)}$ is small, neglects the SOC related to the motion of electron in the $\vec{\rho}$ plane, which in general contributes to the SOC by the boundary conditions.

Below we investigate a InAs $50$~nm-wide core (measured facet-to-facet) surrounded by a $30$~nm InAs$_{1-x}$P$_x$ shell, with $x=0.1$ which allows to neglect strain-induced SOC.\footnote{Lattice 
mismatch and strain field are additional sources of SOC, which are neglected for the present lattice matched heterostructures. We also neglect a SOC contribution resulting from the interface inversion asymmetry related to different atomic termination at opposite interfaces~\cite{Ivchenko,Rossler} which is small. Moreover, additional calculations with slight smearing of the potential at the interface instead of the step like function  did not lead to significantly different results.} 
Furthermore, as shown below, interfacial SOC is enhanced by the easy penetration of envelope functions in low band offset barriers, here only $48$~meV high. 
Simulations have been carried out for a temperature $T=4.2$~K, in the constant electron concentration regime. The parameters adopted are given in Tab.~\ref{Tab:BulkPars}. 
$P$ is assumed to be constant throughout the materials and $E_P(\text{InAs})=2m_0P^2/\hbar^2=21.5$~eV.

\begin{table}
    \begin{tabular}{|c|c|c|c|}
        \hline
        & InAs & InAs$_{0.9}$P$_{0.1}$ \\
        \hline
        $m^*[m_0]$ & 0.0265 & 0.0308  \\
        $E_c[eV]$ & 0.252 & 0.3 \\
        $E_0[eV]$ & 0.42 & 0.5 \\
        $\Delta _0[eV]$ & 0.38 & 0.35 \\
        \hline
    \end{tabular} 
	\caption{Bulk parameters used in calculations~\cite{Vurgaftman2001}.}
	\label{Tab:BulkPars}
\end{table}

\begin{figure}[!ht]
	\begin{center}
		\includegraphics[scale=0.35]{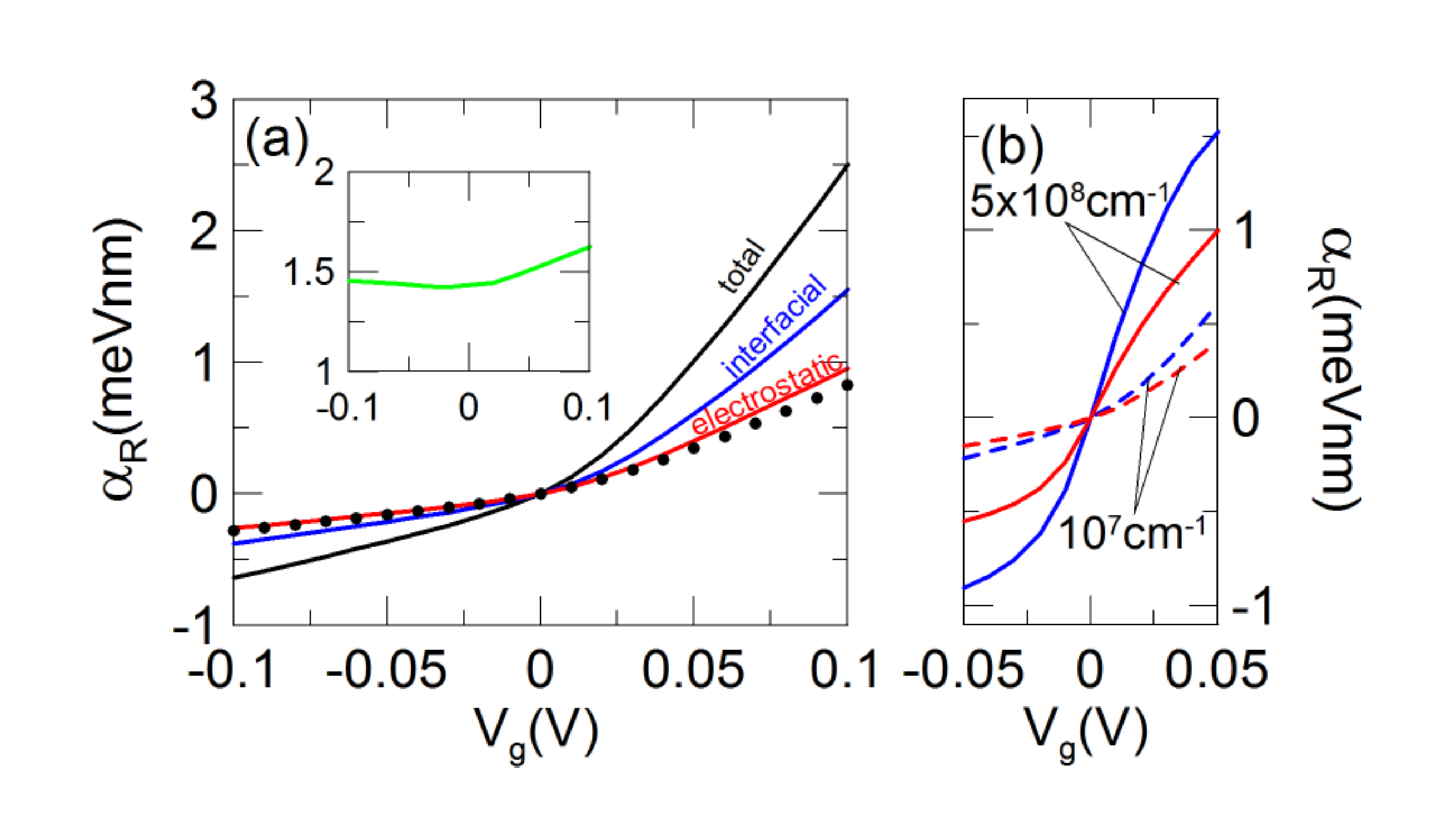}
		\caption{(a) Lines: total, electrostatic and interfacial SOC constants \emph{vs} gate voltage $V_g$, according to labels. Dots: total SOC constant for an equivalent homegeneous InAs 
NW. Inset: ratio between interfacial and electrostatic components, $\alpha^{int}_R/\alpha^{V}_R$. Results for $n_e = 10^7 \mathrm{cm}^{-1}$. See text for structure and material parameters. (b) 
Electrostatic and interfacial SOC constants \emph{vs} $V_g$ around $V_g=0$ showing shooting up of SO couplings for higher electron density. }
		\label{fig2}
	\end{center}
\end{figure}

\begin{figure}[!ht]
	\begin{center}
		\includegraphics[scale=0.5]{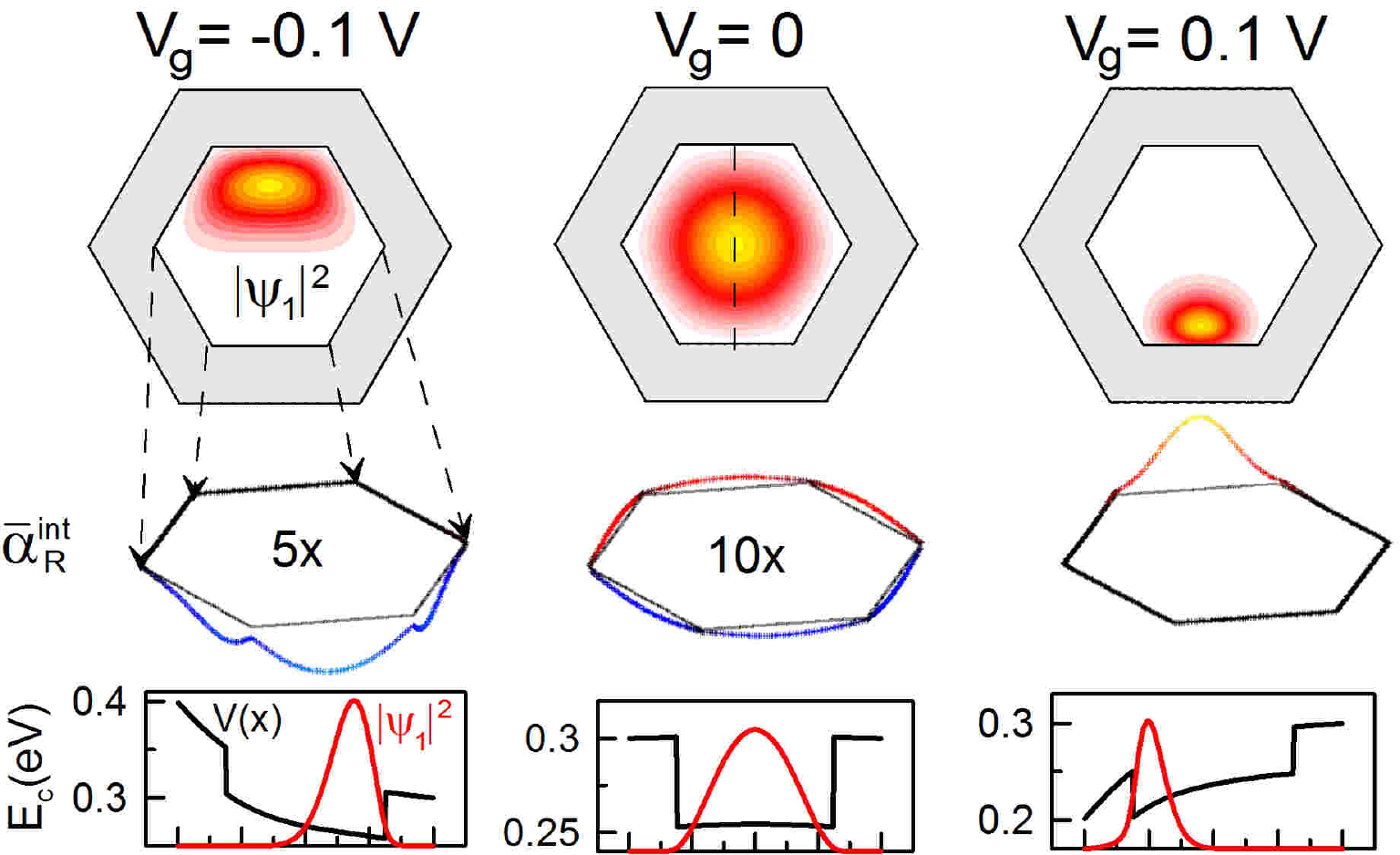}
		\caption{Top row: Square of the ground state envelope function $|\psi_1(x,y)|^2$. Middle row: linear density of the interfacial SOC constants at interfaces. Bottom row: 
self-consistent potential profile (black line) and $|\psi_1(x,y)|^2$ (red line) along the facet-to-facet dashed line marked in the top-middle panel. Results at selected gate voltages 
$V_g=-0.1,0,0.1$~V for the same structure as in Fig.~2.}
		\label{fig3}
	\end{center}
\end{figure}

The calculated SOC coefficients for the InAs/InAsP CSNW of Fig.~\ref{fig1} as a function of the back gate voltage is reported in Fig.~\ref{fig2}(a). 
The SOC constant is trivially zero if $V_g=0$, due to the overall inversion symmetry.
At any finite voltage the inversion symmetry is removed, hence $\alpha_R \ne 0$.
As shown in Fig.~\ref{fig2}(a), the total $\alpha_R$ ensues from two different contributons, namely interfacial and electrostatic, whose magnitude is of the same order. 
It is thus crucial to include both of them in the assessment of SOC in CSNWs.
Note that the electrostatic component almost coincides with the value for an InAs NW with the same geometry, but no overgrown shell\footnote{For the model CSNW under consideration, the gate is attached directly to the NW, so that $\alpha_R$ is an upper bound value, while any dielectric interlayer would decrease $\alpha_R$. For the NW with no shell, the gate has been placed at the same distance from the core as in the case with the shell, in other to compare the two calculations.}.
However, for this specific nanostructure, the largest part of $\alpha_R$ is due to the interfacial contribution, which is $\approx 50\%$ larger than the electrostatic one. 
While the ratio between the two contributions is nearly independent of $V_g$ [see the inset in Fig.~\ref{fig2}(a)], they are both strongly anisotropic with respect to the field direction.
This is due to the different effects on the charge density, as discussed in Ref.~\onlinecite{Wojcik2018}. This effect can also be grasped from the probability distribution reported in Fig.~\ref{fig3} (top and bottom rows).
Indeed, the positive $V_g$ pushes electron states towards the interface opposite to the gate, where the gradient of the self-consistent field is low.
On the other hand, at $V_g<0$ electrons are pulled to the region of the nearest interface with the stronger electric field, additionally strengthened by the electron-electron interaction\cite{Wojcik2018}.

The value of $\alpha^{int} _R$ depends on the penetration of the wave function into the interfaces. As shown in Fig.~\ref{fig3} (middle row) the linear density of interfacial SOC at the interfaces 
$\bar{\alpha}^{int}_R = \psi_1(\vec{\rho}) \alpha^{int}_{R}(\vec{\rho}) \psi_1(\vec{\rho})$ is finite almost everywhere, but it has opposite sign at opposite facets.\footnote{The sign of the SOC 
linear density on each facet is decided by the choice of the reference frame. Since $\hat{\alpha}_x$ appears combined with the spin matrices in the Hamiltonian (\ref{eq:3DH}), the overall sign of the 
SOC term is unaffected by a change in the reference frame.} For a centro-symmetric system ($V_g=0$) the overall value is zero, since opposite contributions cancel out exactly. We 
stress a remarkable difference between CSNWs and analogous planar structures. In a \textit{planar} asymmetric quantum well, for example, $\alpha_R\ne 0$. In a CSNW with an embedded quantum well, 
however, the overall symmetry is recovered even if each facet of the quantum well is individually asymmetric. Therefore, opposite segments have opposite Rashba 
contributions and compensate. However, any asymmetric gate potential unbalances opposite contributions, the total effect being related to the amount of envelope function at the interface. 

\begin{figure}[!ht]
	\begin{center}
		\includegraphics[scale=0.35]{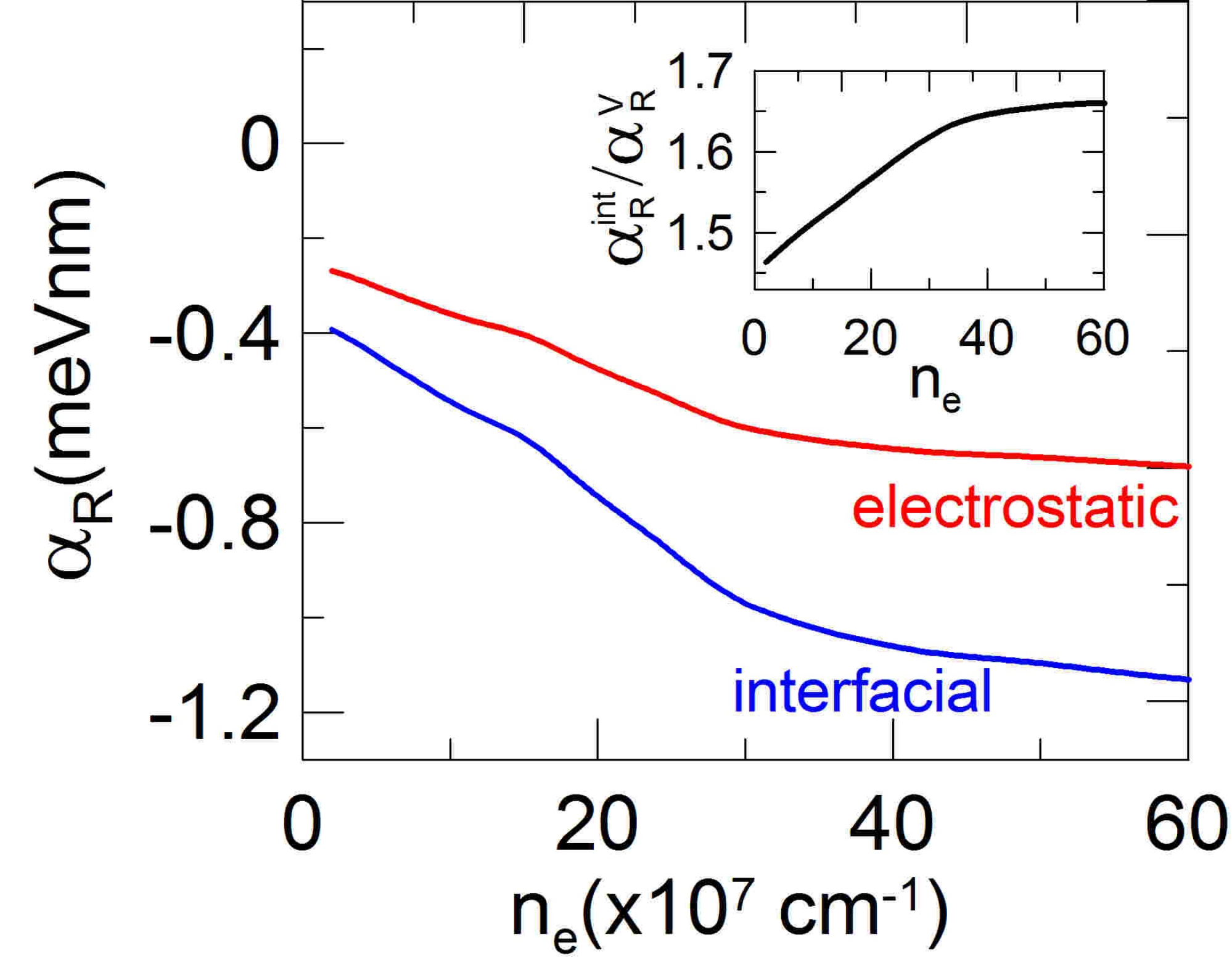}
		\caption{Electrostatic $\alpha^{V}_R$ and interfacial $\alpha^{int} _R$ contributions of SOC constant \textit{vs} electron density $n_e$. Results for $V_g=-0.1$~V. Inset: ratio 
$\alpha^{int}_R/\alpha^{V}_R$ \textit{vs} $n_e$.}
		\label{fig4}
	\end{center}
\end{figure}

Note the almost linear increase of $\alpha_R$ with $V_g$. This behaviour is observed in a relatively small charge density regime: the average Coulomb energy is small, most of the charge is located in 
the core, and it is relatively rigid to an applied transverse electric field. At larger densities, however, charge moves at the interfaces to minimize Coulomb interaction~\cite{Bertoni2011}, with 
negligible tunneling energy between opposite facets. In this regime, the symmetric charge density distribution is unstable and it is easily distorted by an electric field~\cite{Wojcik2018}. 
Accordingly, SOC constant shoots around $V_g=0$ as soon as the gate is switched on - see Fig.~\ref{fig2}(b).

\begin{figure}[!ht]
	\begin{center}
		\includegraphics[scale=0.45]{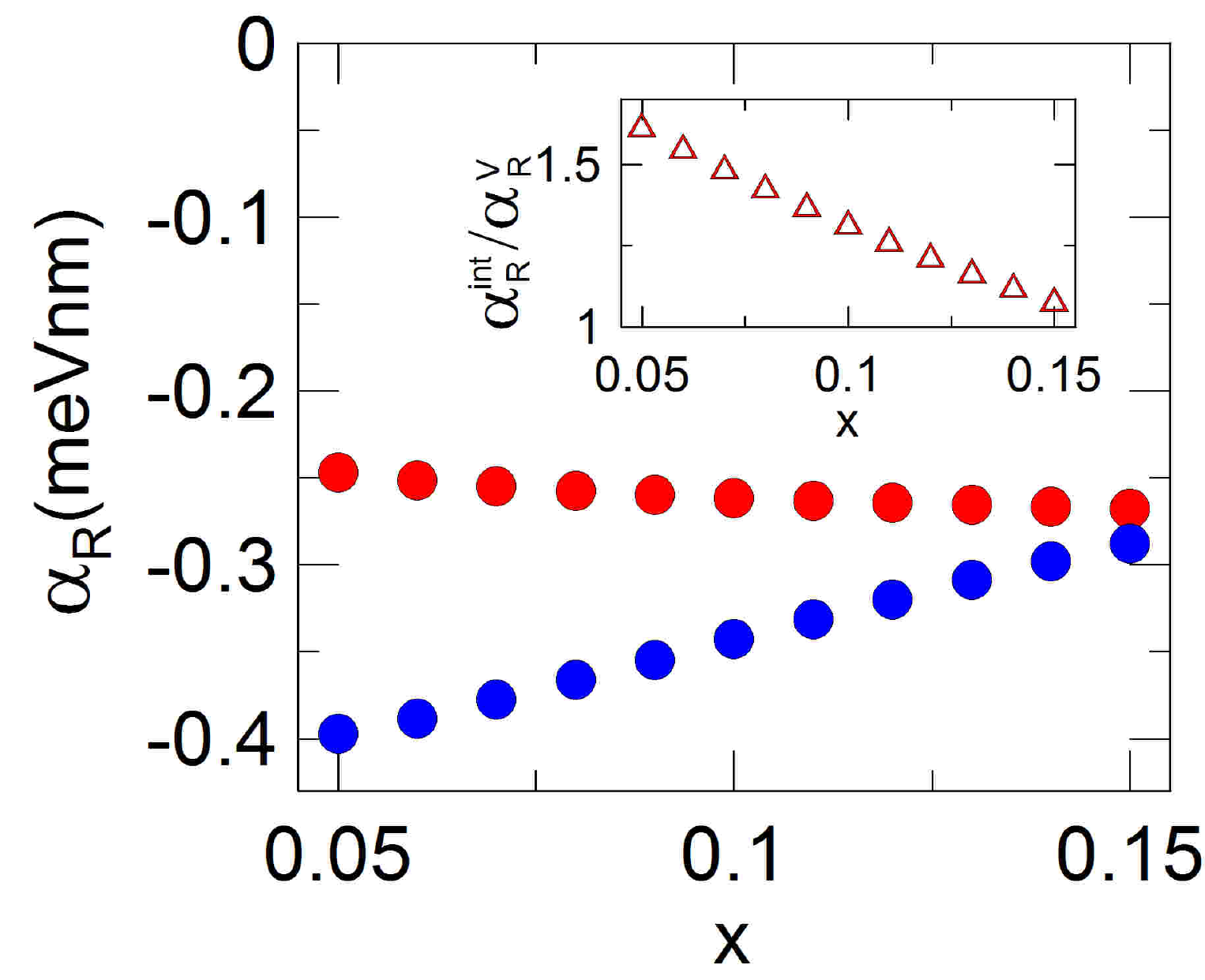}
		\caption{The interfacial $\alpha^{int}_R$ (blue circles) and  electrostatic $\alpha ^{V}_R$ (red circles) SOC constants \textit{vs} InAs$_{1-x}$P$_x$ alloy composition, $x$. Inset: $\alpha^{int}_R/\alpha^{V}_R$ \textit{vs} $x$. Results for $V_g=-0.1$~V and $n_e=10^{7}$~cm$^{-1}$.}
		\label{fig5}
	\end{center}
\end{figure}

As we show in Fig.~\ref{fig4}, both SOC components substantially increase in intensity with charge density, while their ratio is weakly affected, it being rather \textit{a property of the 
nano-material} (band parameters and band offset). This is explicitly shown in Fig.~\ref{fig5}, where the two contributions are plot \textit{vs} the stechiometric fraction $x$.
At low $x$, penetration is very large, and the interfacial effect is dominant (of course for $x=0$ the heterostructure is an homogeneous NW with a larger diameter), while as $x=0.15$ the two 
contributions are comparable, as also shown in the inset.

To summarize, we have shown that Rashba SOC in CSNWs is increased by the effect of the radial heterointerface, and its control via external metallic gates may be highly improved by this interfacial effect.
Although we did not attempt to optimize $\alpha_R$ in the many parameter space allowed by CSNWs, our results suggest that a general strategy to enhance SOC in CSNWs relies on a modification of the 
compositional structure exploiting asymmetric penetration of the wave function into the shell layer.

This work was partially supported by the AGH UST statutory tasks No. 11.11.220.01/2 within subsidy of the Ministry of Science and Higher Education and in part by PL-Grid Infrastructure. P.W. was 
supported by National Science Centre, Poland (NCN) according to decision 2017/26/D/ST3/00109. 

%

\end{document}